\documentclass[
superscriptaddress,
twocolumn,
amsmath,amssymb,
aps,
prb,
floatfix,
showkeys,
reprint,
]{revtex4-2}

\usepackage{graphicx}
\usepackage{dcolumn}
\usepackage{bm}
\usepackage{amsmath}
\usepackage{amssymb}
\usepackage{multirow}
\usepackage{xcolor}
\usepackage{booktabs}
\usepackage{tabularx}
\usepackage[english]{babel}
\usepackage[utf8]{inputenc}
\usepackage{enumerate}
\usepackage{newunicodechar}
\newunicodechar{ﬁ}{fi}
\newunicodechar{ﬀ}{ff}
\usepackage[colorlinks, citecolor=blue, linkcolor=blue]{hyperref}
\usepackage{gensymb}
\usepackage{float}

\usepackage{lipsum}

\graphicspath{{figs/}}

\begin{document}
    \title{First-principles investigation of magnetic exchange force microscopy on \\ adatoms adsorbed on an antiferromagnetic surface}

\author{Soumyajyoti Haldar}
\email[Corresponding author: ]{haldar@physik.uni-kiel.de}
\affiliation{Institute of Theoretical Physics and Astrophysics, University of Kiel, Leibnizstrasse 15, 24098 Kiel, Germany}

\author{Stefan Heinze}
\affiliation{Institute of Theoretical Physics and Astrophysics, University of Kiel, Leibnizstrasse 15, 24098 Kiel, Germany}
\affiliation{Kiel Nano, Surface, and Interface Science (KiNSIS),  University of Kiel, 24118 Kiel, Germany}

\date{\today}

\begin{abstract}
Using density functional theory (DFT), we calculate the magnetic short-ranged exchange forces between a magnetic tip and an adatom adsorbed on the antiferromagnetic Mn monolayer on the W(110) surface [Mn/W(110)]. These exchange forces can be measured in magnetic exchange force microscopy allowing
atomic-scale imaging of spin structures on insulating and conducting surfaces.   
We consider two types of $3d$ transition-metal atoms with intrinsic magnetic moments: Co and Mn and Ir as an example of a $5d$ transition-metal atom 
exhibiting an induced magnetic moment on Mn/W(110). The tips are modeled by Fe pyramids and terminated either with an Fe or a Mn apex atom. 
From our total energy DFT calculations for a parallel and antiparallel alignment between tip and adatom magnetic moments we obtain the exchange energy $E_{\rm ex}(d)$ as a function 
of tip-adatom distance $d$. The exchange forces, $F_{\rm ex}(d)$, are calculated based on the 
Hellmann-Feynman theorem. We show that structural relaxations of
tip and sample due to their interaction need to be taken into account. Due to the exchange
interaction the relaxations depend on the alignment between tip and adatom magnetization
-- an effect which will affect the tunneling magnetoresistance that can be measured
by a scanning tunneling microscope. 
A maximum in the exchange energy and force curves is 
obtained for magnetic adatoms at tip-adatom 
separations of about 3 to 4~{\AA}.
The exchange forces with an Fe terminated tip reach a maximum value of up to 0.2~nN and 0.6~nN 
for Co and Mn adatoms, respectively, and prefer an antiferromagnetic coupling. 
Surprisingly, we also find an exchange force of up to 
0.2~nN for Ir adatoms. We analyze the exchange interaction between tip and adatom based on
the spin-polarized electronic structure of the coupled system. A competition occurs between 
long-range Zener-type indirect double exchange favoring ferromagnetic coupling and short-range direct 
$d-d$ antiferromagnetic exchange. For the Ir adatom
the interaction can be explained from the spin-dependent
hybridization with the tip apex atom.
Our results show that magnetic adatoms on Mn/W(110) are a promising system 
to study exchange forces at the single-atom level
via magnetic exchange force microscopy.

\end{abstract}

\maketitle

\section{Introduction}
Non-collinear magnetic structures, such as chiral domain walls ~\cite{Heide2008,Ryu2013,Emori2013}, skyrmions~\cite{Bogdanov1989,Bogdanov:2001aa,Heinze2011,Romming2013,Herve2018,Back2020,Paul2020}, and spin spirals~\cite{Ferriani2008,Phark2014} 
are being extensively studied for future
spintronic applications due to their %
exceptional dynamical and transport properties~\cite{Fert2013,Nagaosa2013}.  
A prominent example for a material system exhibiting a cycloidal spin-spiral ground state is a monolayer of Mn grown on the W(110) surface [Mn/W(110)]. Here the cycloidal spin-spiral state is driven by the Dzyaloshinskii-Moriya interaction and propagates along the [1$\overline{1}$0] direction with an angle of $\sim$ 173{\degree} between the magnetic moment of the adjacent rows~\cite{Bode2007}. Single magnetic adatoms and nanostructures on 
surfaces are of 
interest for
research in the field of magnetic data storage~\cite{Gambardella2003,Lounis2007,Enders2010,Donati2016,Hermenau2017}. 
Using the non-collinear magnetic state of Mn/W(110)
the spin orientation of an adsorbed magnetic adatom can be controlled by the local exchange coupling 
without the presence of an external magnetic 
field~\cite{Serrate2010,Serrate2016,Haldar2018stm}.   
This allows a direct measurement of exchange interaction for single atoms adsorbed on a non-collinear magnetic surface and an understanding of their origin which is of fundamental interest and can be important for potential applications.

In the field of nanomagnetism and spintronics, spin-polarized scanning tunneling microscopy (SP-STM) and spectroscopy (SP-STS) play an important role 
to probe the properties of single magnetic atoms, molecules, or ultrathin films with atomic-scale resolution \cite{Bode2003,Wiesendanger2009Rev}. 
The development of these techniques has led to novel methods of magnetometry at the single atom level~\cite{Meier82,Zhou2010,Heinrich2004,Hirjibehedin1199,Wiesendanger2011}. SP-STM can also be used to measure the transport properties of magnetic adatoms as a function of spin direction~\cite{Wiesendanger1990,Wulfhekel1999,Bode2003},
tip-adatom distance \cite{Ziegler2011,Lazo2012},
or can be used to control the spin directions of an adatom adsorbed on a noncollinear magnetic surface~~\cite{Serrate2010,Serrate2016}. However, SP-STM and SP-STS are limited to electrically conducting samples such as magnetic metals or magnetic semiconductors and not suitable to measure the exchange force between two atomic magnetic moments. Furthermore, the tunneling current is sensitive to subtle electronic structure
effects leading to the tunneling anisotropic magnetoresistance~\cite{Bode2002,Neel2013} 
and the non-collinear magnetoresistance \cite{Hanneken2015,Crum2015,Perini2019} which can severely affect the measured signal and can make its interpretation challenging.  

Complementary to the spin-sensitive tunneling methods, magnetic exchange force microscopy (MExFM) is an approach based on non-contact atomic force microscopy %
which allows to measure the local exchange force between a magnetic tip and a magnetic surface~\cite{Kaiser2007,Schmidt2009,Schmidt2011}. 
A force-microscopy-based detection of magnetic interaction has %
advantages over the tunneling current-based techniques. First of all, MExFM technique %
can be used for both conducting~\cite{Schmidt2009,Grenz2017} and insulating magnetic systems~\cite{Kaiser2007,Pielmeier2013}. 
Furthermore, the exchange energy and force can be measured quantitatively using magnetic exchange force spectroscopy~\cite{Schmidt2011,Hauptmann2017}. 
Since the exchange interaction between two magnetic moments 
is at the heart of magnetic phenomena its quantification is of fundamental interest.
In combination with first-principles calculations, this allows to reveal
various exchange mechanisms~\cite{Tao2009} and
characteristic features due to chemical functionalization of the tip~\cite{Lazo2008,Lazo2011}.

Recently, it has been demonstrated that one can combine SP-STM and MExFM measurements in one instrument,
which allows to obtain the spin-polarized (SP) current and the exchange (EX) forces simultaneously.
This technique has therefore been coined SPEX imaging~\cite{Hauptmann2017}.
A simultaneous measurement of the local exchange force and the spin-polarized tunneling current between a magnetic tip and a surface with a non-collinear spin structure has been obtained~\cite{Hauptmann2017,Hauptmann2019}. 
However, such investigations for adatoms adsorbed on a non-collinear magnetic surface are missing both theoretically and experimentally.  

Here, we aim to obtain a better understanding of how MExFM technique equipped with a magnetic tip can be used in non-contact mode 
to measure the short-ranged magnetic exchange force on magnetic adatoms.
We investigate the distance dependence of the magnetic exchange force, its size and sign between a magnetic tip and an adatom adsorbed on 
the surface of Mn/W(110) using density functional theory (DFT). In this investigation, we have chosen three different adatoms: Co, Mn, and Ir. While Co and Mn both exhibit intrinsic magnetic
moments, Ir obtains only a small induced magnetic moment due to the hybridization with
the Mn surface atoms. 
In an experiment the magnetic tip often picks up an atom from the surface due to intentional or unintentional collisions. Therefore, for the magnetic Fe tip, we have considered Fe base atoms and a termination with either an Fe or a Mn apex atom
(see sketch in Fig.~\ref{fig:schematics}).

We calculate the exchange energy and force as a function of tip-adatom distance
taking structural relaxations due to tip-adatom interaction into account. The
structural relaxations depend on the spin alignment of tip and adatom and
have a significant impact on the exchange forces. They also modify the tunneling magnetoresistance (TMR) that can be measured simultaneously in SPEX imaging. For 
the Mn adatom the change of the TMR can be up to 15\% at the closest distances.
Based on an analysis of spin-resolved charge density difference plots we conclude
that the exchange interaction between the Fe tip and Co and Mn adatoms results
from a competition of short-range direct $d-d$ antiferromagnetic exchange and
long-range indirect Zener-type ferromagnetic coupling.

The structure of the paper is as follows. First, we discuss the applied computational methods in Sec.~\ref{sec:computation}. The main part of the paper (Sec.~\ref{sec:results}) contains the results and discussions for the investigated magnetic exchange energy and forces between magnetic tip and the three adatoms (Co, Mn, and Ir) adsorbed on Mn/W(110) surface. Finally, we summarize our main conclusions in Sec.~\ref{sec:conclusion}.

\section{Computational details}
\label{sec:computation}
\subsection{Computational methods}
\begin{figure}[htbp!]
	\centering	\includegraphics[width=0.9\linewidth]{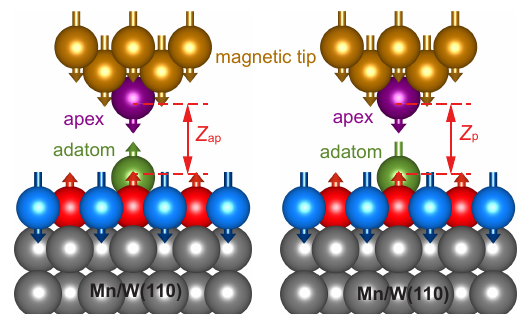}
	\caption{Schematic illustration of a ferromagnetic tip approaching a 
    single magnetic atom adsorbed on the Mn/W(110) surface in an MExFM
    experiment. 
 Left [right] panel shows the antiparallel (ap) [parallel (p)] alignment of the magnetic moments of the
 tip apex (purple) and absorbed adatom (green). The distance between
 the tip apex atom and the adatom differs in general for the
 antiparallel, $z_{\rm ap}$, and parallel alignment, $z_{\rm p}$.
 W atoms are given
 by grey spheres and Mn atoms with opposite magnetic moment 
 directions (arrows) are shown as blue and red spheres.
 }
	\label{fig:schematics}
\end{figure}
In this work we have used DFT
calculations within the projector augmented wave method (PAW)~\cite{blo,blo1} as implemented in the \textsc{vasp} code~\cite{vasp1,vasp2}. For the exchange correlation we have used the generalized gradient approximation (GGA) of Perdew-Burke-Ernzerhof (PBE)~\cite{PBE,PBE2}. The combined system consisting of both the magnetic tip and the sample was modelled in a supercell geometry [see Fig.~\ref{fig:schematics}]. We modelled the Mn/W(110) surface using a symmetric slab with 5 W layers and 1 layer of Mn on each side. The local magnetic order of the Mn/W(110) surface was chosen as collinear antiferromagnetic (AFM). This is a good approximation  
due to the long periodicity of the spin spiral ground state which
leads to angles of about 173$^\circ$ between the magnetic moments 
of Mn atoms in adjacent rows \cite{Bode2007}.
We designed the substrate using a c(6$\times$6) AFM surface unit cell with GGA lattice constant of W (3.17 {\AA}), which is in good agreement with the experimental lattice constant of 3.165 {\AA}. The respective adatom is centered in the hollow-site position within the unit cell on both sides of the slab. This position is the energetically most stable adsorption site~\cite{Gutzeit2020}. The tip was modeled by a 14-atom pyramid consisting of ferromagnetic Fe atom in a bcc(001) orientation [see Fig.~\ref{fig:schematics}] as in Ref.~\cite{Lazo2011}. 

We have relaxed the tip and the surface structure individually before considering the tip-sample interaction. For isolated tips (without any tip-sample interaction) the apex atom and the four atoms of the adjacent layer have been relaxed in out-of-plane direction only. The in-plane 
inter-atomic distances between the base atoms are kept constant at the GGA lattice constant of Fe (2.85 {\AA}). In our calculations, we have also considered the effect of the structural relaxations due to the tip-sample interactions. We have used a 5 $\times$ 5 $\times$ 1 $\overline{\Gamma}$ centered $k$-grid mesh~\cite{Monkhorst} for the Brillouin zone integration and used a 450 eV energy for the plane wave basis set cutoff. The force tolerance for the structural minimization were used as 0.01eV/{\AA}. We calculate the force by using the Hellmann-
Feynman theorem~\cite{Feynman1939}. 

\subsection{Exchange energy \& force}
We have considered the influence of two different tip terminations for the Fe-based tips
by choosing either an Fe apex atom or a Mn apex atom.
For a Mn-terminated tip, antiferromagnetic coupling occurs between the Mn apex atom
and the Fe base atoms, while the Fe-terminated tip is ferromagnetic. 
For every tip-sample separation $d$ 
-- defined as the distance between the centers of the tip apex atom and the adatom underneath without structural relaxations due to tip-sample interaction -- 
we considered a parallel ($p$) and an antiparallel ($ap$) alignment of the magnetic moments between the tip apex atom and the adsorbed adatom. We calculate the magnetic exchange energy, $E_{\mathrm{ex}}(d)$, from the total energies of the $ap$- and $p$- configurations:
\begin{equation}
    E_{\mathrm{ex}}(d)=E_{\mathrm{ap}}(d)-E_{\mathrm{p}}(d) \;,
\end{equation}
where $E_{\mathrm{ap}}(d)$ and $E_{\mathrm{p}}(d)$ denotes the total energies of the $ap$- and $p$-configurations, respectively. Hence, $E_{\mathrm{ex}}(d) > 0$ and $E_{\mathrm{ex}}(d)<0$ indicate ferromagnetic and antiferromagnetic coupling, respectively. Similar to the magnetic exchange energy, we define the magnetic exchange force as follows:
\begin{equation}
    F_{\mathrm{ex}}(d)=F_{\mathrm{ap}}(d)-F_{\mathrm{p}}(d) \;,
\end{equation}
here $F_{\mathrm{ap}}(d)$ and $F_{\mathrm{p}}(d)$ are the total forces on the tip along the $z$ direction of the Mn/W(110) (i.e., the [110] direction) for the $ap$- and $p$-configuration, respectively. 

The tip-sample distance $d$ still holds true when we consider the relaxation as we fixed the uppermost layer of the tips and added the unrelaxed apex atom positions of the isolated tips in order to calculate the ideal $d$ of the tip apex atom. This definition mimic the experimental situation where only the relative displacement of the tip body can be measured and the exact distance between the tip apex atom and underlying sample is unknown. 
From our DFT calculations we can determine the actual distance between
the tip apex atom and the adatom. Due to the exchange interaction, the distance 
$z_{\rm p}(d)$ for the $p$-alignment differs from $z_{\rm ap}(d)$ for the $ap$-alignment 
(cf.~Fig.~\ref{fig:schematics}).

\section{Results and Discussion}
\label{sec:results}
\subsection{Co adatom on Mn/W(110)}

\begin{figure}[htbp!]
	\centering
	\includegraphics[width=0.9\linewidth]{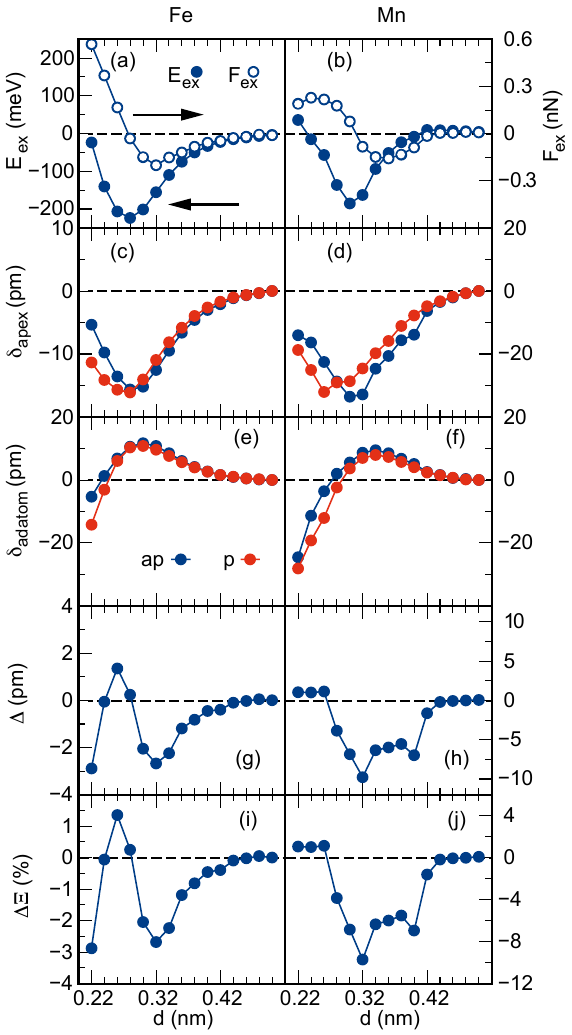}
	\caption{(a,b) Exchange energy (solid dots) and exchange force (hollow circles) as a function of distance $d$ between an Fe tip with an Fe (left panels) and a Mn (right panels) apex atom and a Co adatom adsorbed on the
    Mn/W(110) surface. (c,d) Structural relaxations of the tip apex atom, $\mathrm{\delta_{apex}}$, and (e,f) of the adatom, $\mathrm{\delta_{adatom}}$, for p (red dots) and ap (blue dots) configurations. The positive relaxation values for the apex atom of the tip indicate downward (toward the substrate) structural relaxation, while the positive values for the Co adatom denote upward (toward the tip) structural relaxation. (g,h) Relaxation difference $\Delta (d)$ between $p$ and $ap$ configuration. (i,j) Calculated TMR difference $\Delta \Xi(d)=\Xi(d)-\Xi_0$. Note the different values
    on the left and right $y$-axis for all panels.}
	\label{fig:SPEX_Compare_on_Co}
\end{figure}

We start our discussion with the magnetic exchange interaction between a magnetic tip and a Co adatom adsorbed on Mn/W(110) which includes the structural relaxations of tip and sample atoms due to their interaction (Fig.~\ref{fig:SPEX_Compare_on_Co}). This case is of particular interest for our theoretical MExFM study
since SP-STM experiments have already been performed for Co adatoms on Mn/W(110) \cite{Serrate2010,Serrate2016}. The Co adatom couples ferromagnetically with
respect to the nearest-neighbor Mn surface atoms and exhibits a magnetic moment
of about 1.5~$\mu_{\rm B}$ \cite{Serrate2010,Gutzeit2020}.
In an experiment, the magnetic tip can often pick up an atom from the surface either due to intentional or due to unintentional collisions. Therefore, 
we have used Fe-based tips with an Fe and with a Mn apex atom (left and right panels of Fig.~\ref{fig:SPEX_Compare_on_Co}). 

The calculated magnetic exchange energies and forces are shown in Figs.~\ref{fig:SPEX_Compare_on_Co}(a,b)
for the Fe and Mn tip apex atom, respectively. The magnetic exchange energy (solid blue circles) for the Fe-apex atom tip and Mn-apex atom tip have a similar trend. They are negligible at large distances ($d>$ 0.5 nm) and drop as the magnetic tip approaches towards the adsorbed adatom. From the exchange energies, Fig.~\ref{fig:SPEX_Compare_on_Co}(a), we see a clear regime of antiferromagnetic coupling ($E_\mathrm{ex} < 0$) for the Fe-apex atom tip for all the distances. However, for the tip with a Mn apex atom, Fig.~\ref{fig:SPEX_Compare_on_Co}(b), a strong antiferromagnetic coupling at smaller distances changes to a weak ferromagnetic coupling ($E_\mathrm{ex} > 0$) for larger distances ($d> 0.42$ nm). At small distances close
to the maximum absolute value of $E_\mathrm{ex}$, the strength of the antiferromagnetic couplings are quite 
similar for both types of tips.
The strongest coupling is observed at $d\approx 0.28$~nm and $d\approx 0.30$~nm for the Fe-apex and Mn-apex tip, respectively. 

The resulting exchange forces (hollow blue circles in Figs.~\ref{fig:SPEX_Compare_on_Co}(a,b)) for both 
tips behave quite similar as well. For a distance 
down to about $d=0.34$ nm the forces are negative and increase in magnitude. In both force curves we find a 
minimum with a value of $F_{\rm ex} \approx -0.2$~nN which is shifted to larger distances for the 
Mn-terminated tip. However, the exchange force minimum occurs at smaller distances for the Fe-terminated tip
while the exchange force
rises already to positive values for the Mn-terminated tip at about $0.3$~nm.

In order to understand the impact of structural relaxations on the observed exchange interaction, 
we have investigated the displacement of the tip apex ($\delta_\mathrm{apex}$) and the adsorbed adatom 
($\delta_\mathrm{adatom}$). Note, that these depend on the magnetic configuration, i.e.~$p$- or $ap$-alignment
between tip apex and adatom magnetic moment.
Fig.~\ref{fig:SPEX_Compare_on_Co}(c-f)
summarize the structural relaxations of the tip apex atom and the probed Co adatom on Mn/W(110). The displacement of the surface Mn layer is below 2 pm and is not shown. At a tip sample distance of $d>0.5$ nm the relaxations become negligible. As the tip approaches toward the surface, the attractive force increases significantly and the distance between the tip-apex atom and the adsorbed adatom decreases. However, at a small distance, close to the contact regime, the repulsive interaction due to overlapping electron shells starts to dominate and the distance between the tip and the adatom increases again. 

For all cases, the displacement of the tip-apex atom is more strongly affected compared to the adatom due to its lower coordination number. The displacement of the Co adatom reaches up to $\sim 10$ pm whereas the displacement of the Fe and the Mn apex atom are at most $\sim 15$ pm and $\sim 30$ pm, respectively. The exchange force on the apex atom is negative at larger distances ($d>0.28$~nm). Hence the displacement of tip-apex atom and the adatom is larger in the antiparallel alignment because of an enhanced attractive interaction. The difference between the relaxations in the $ap$- and $p$-configurations is relatively small for the Fe apex atom. For Mn-terminated tip, the relaxations are considerably different in the two magnetic configurations, in particular, at small distances for $d<0.4$~nm.     
The exchange energy drops more slowly and accordingly,
the magnitude of the exchange force starts to decrease.
Note, that structural relaxations of similar magnitude have been obtained for simulations of STM experiments on
nonmagnetic surfaces~\cite{Hofer2001,Blanco2004} and adatoms~\cite{Ternes2011} as well as for antiferromagnetic surfaces~\cite{Lazo2008,Lazo2011} and magnetic adatoms on a ferromagnetic surface \cite{Lazo2012}.

The calculated relaxation difference $\Delta(d)=z_\mathrm{ap}(d) - z_\mathrm{p}(d)$ is shown in Figs.~\ref{fig:SPEX_Compare_on_Co}(g,h). 
For the tip with an Fe apex atom, a maximum absolute value of 3~pm is obtained at the minimum 
of the exchange force. For the Mn-terminated tip, the effect is much more pronounced with 
a maximum value of up to 10~pm. This shows that structural relaxations have a much larger
impact on tips with a Mn apex atom.
Interestingly, $\Delta(d)$ can also be obtained from the SP-STM experiments by measuring the tunneling magnetoresistance as a function of distance as described below following 
Ref.~\cite{Lazo2012}. 

The magnetic exchange dependent relaxation difference between the magnetic tip apex atom and the adsorbed adatom, $\Delta(d)$, can be experimentally quantified by measuring the tunneling magnetoresistance $\Xi$, which is defined as
\begin{equation}
    \Xi(d)=\frac{I_\mathrm{p}(z_\mathrm{p}(d)) - I_{\mathrm{ap}}(z_{\mathrm{ap}}(d))}{I_\mathrm{p}(z_\mathrm{p}(d)) + I_{\mathrm{ap}}(z_{\mathrm{ap}}(d))}\;.
\end{equation}
Here $I_\mathrm{p}$ and $I_{\mathrm{ap}}$ represent the tunneling currents for parallel and antiparallel spin alignment, respectively. $z_{\mathrm{p}}(d)$ and $z_{\mathrm{ap}}(d)$ are defined as the actual distance between the tip apex atom and the adatom for the parallel and antiparallel spin alignment, respectively (cf.~Fig.~\ref{fig:schematics}). The distance between the apex and the adsorbed adatom without the relaxation due to tip-sample interaction 
is denoted by $d$. 

In an SP-STM experiment, the tunneling current can be written as $I_{\mathrm{p,ap}}(d) = (I_\mathrm{0} \pm I_{\mathrm{sp}}) \, \mathrm{exp}(-2\kappa d)$ where $I_\mathrm{0}$ and $I_{\mathrm{sp}}$ are the non-spin-polarized and the spin-polarized part of tunneling current, respectively, and $\kappa=\sqrt{2m \phi/\hbar^2}$ is the decay 
constant \cite{Bode2003,Wiesendanger2009Rev,Wortmann2001}.
Therefore, the tunneling magnetoresistance (TMR) can be written as
\begin{equation}
\label{eq:xi}
    \Xi(d)=\Xi_\mathrm{0} + \frac{1-\mathrm{exp}[-2\kappa\Delta(d)]}{1+\mathrm{exp}[-2\kappa\Delta(d)]}\;.
\end{equation}
Here, $\Xi_\mathrm{0} = I_\mathrm{sp}/I_\mathrm{0}$ is the TMR at large
separations where structural relaxations can be neglected.
Therefore, using Eq.~(\ref{eq:xi}), $\Delta(d)$ reads as
\begin{equation}
\label{eq:delta}
    \Delta (d)=-\frac{1}{2\kappa}\mathrm{ln} \left( \frac{1 - \Xi(d) + \Xi_\mathrm{0}}{1 + \Xi(d) - \Xi_\mathrm{0}}\right) \;.
\end{equation}
$\Delta(d)$ can be directly calculated based on Eq.~(\ref{eq:delta}) from an experimental measurement of the distance-dependent TMR $\Xi(d)$. $\kappa$ is determined from the exponential decay of the tunneling current with distance. 

In Fig.~\ref{fig:SPEX_Compare_on_Co}(i,j) we show the change of the TMR calculated
based on Eq.~(\ref{eq:xi}) using the DFT values of $\Delta(d)$ and assuming a value
of $\kappa=1$~{\AA}$^{-1}$. It is apparent that the TMR follows the exchange induced
structural difference. The TMR obtained for the Co adatom changes by up to about 3\% 
and 8\% depending on whether the Fe tip is terminated with an Fe or a Mn apex atom,
respectively.

Note, that values of $\Delta(d)$ obtained in the reverse way, i.e.~using Eq.~(\ref{eq:delta}),
from experimental TMR
data obtained via SP-STM measurements on Cr and Co adatoms on a ferromagnetic surface 
are in good agreement with DFT calculations \cite{Lazo2012}.
However, in those experiments it was not possible to measure the exchange forces
simultaneously. A correlation of exchange forces and relaxations due to exchange
interactions as proposed here would be feasible using SPEX measurements on magnetic adatoms.

 \begin{figure}[htbp!]
	\centering
	\includegraphics[width=0.8\linewidth]{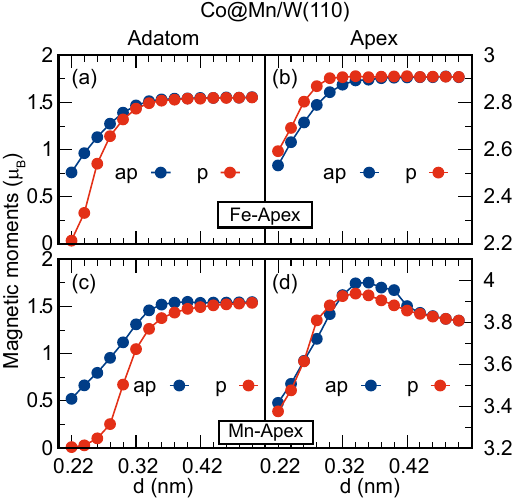}
	\caption{Distance dependence of the magnetic moments as the Fe tip 
     with an Fe or Mn apex atom approaches
     the Co adatom on Mn/W(110). (a,b) magnetic moments of the Co adatom
     and the Fe-apex atom and (c,d) the Co adatom and the Mn-apex atom. Antiparallel (ap) and parallel (p) configurations are shown in blue and red, respectively. Absolute values of the magnetic moments have been plotted to facilitate comparison between both magnetic alignments. }
	\label{fig:magnetic_moment_Co}
\end{figure}

The tip-sample interaction leads to a change of the magnetic moments of the
apex atom and the adatom. This effect is shown in Fig.~\ref{fig:magnetic_moment_Co} for 
the Fe-tip with an Fe or Mn apex atom and the adsorbed Co adatom on Mn/W(110). 
This provides additional insight to the effect of structural relaxations. 
In general, the magnetic moments decrease as
the tip apex atom approaches the Co adatom due to the
increased hybridization between their orbitals. 
For Fe-terminated tip-adatom interaction, we observe that the Fe and Co moments
are nearly constant down to about 0.32~nm. For smaller separations, we find
a decrease in absolute value of magnetic moment for $d<0.32$~nm [Fig.~\ref{fig:magnetic_moment_Co}(a-b)] for 
both $ap$- and $p$-configuration. 
At the smallest considered distance of 0.22~nm the Co moment drops from
its initial value of about 1.5~$\mu_{\rm B}$ to 
0.75~$\mu_{\rm B}$ in the $ap$-alignment and is even quenched in the 
$p$-alignment. 

For the Mn-terminated tip the drop of the magnetic moments is similar at small
distances, however, there is also a subtle difference visible in
Fig.~\ref{fig:magnetic_moment_Co}(c-d). The DFT calculation shows a maximum in the magnetic moment of the Mn apex atom around $d=0.35$ nm for both $ap$- and 
$p$-configuration. 
This feature is related to the comparatively large structural relaxation of the Mn apex atom (see Fig.~\ref{fig:SPEX_Compare_on_Co}) as the tip approaches the surface. The maximum in the structural relaxation curve of the Mn-apex atom signals
a signiﬁcant displacement of the apex from the tip base atoms. As a consequence, the orbitals of the Mn tip apex atom are less hybridized with the orbital of the tip base atoms and its magnetic moment increases before being reduced by the orbital interaction with the Co adatom as the tip approaches the adsorbed Co adatom on the Mn/W(110) surface.
\begin{figure}[htbp!]
	\centering
	\includegraphics[width=1\linewidth]{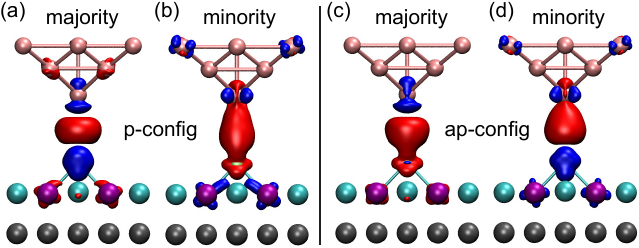}
	\caption{Spin-resolved charge density difference plots along the [1$\overline{1}$0] direction for the interaction of an Fe-tip with a Fe apex atom
    and the Co adatom on Mn/W(110) at a tip sample distance of $d=0.36$~nm.
    (a,b) Majority and minority spin channel, respectively, for the $p$-alignment
    of the magnetic moments of Fe tip and Co adatom. (c,d) as (a,b) for the
    $ap$-alignment. Red and blue isosurfaces denotes charge accumulation and depletion, respectively.}

	\label{fig:charge_density_Co}
\end{figure}

We use spin-resolved charge-density difference (CDD)
plots in order to gain further insight into the electronic and magnetic exchange interaction between the magnetic tip and the adsorbed adatom. The charge density difference, $\Delta \rho_\sigma$, for spin channel $\sigma$ is calculated using 
\begin{equation}
    \Delta \rho_\sigma=\rho^{\mathrm{tip}+\mathrm{sample}}_\sigma - \rho^{\mathrm{tip}}_\sigma - \rho^{\mathrm{sample}}_\sigma \;,
\end{equation}
where $\rho^{\mathrm{tip}+\mathrm{sample}}_\sigma$ denotes  the charge density of the coupled system, $\rho^{\mathrm{tip}}_\sigma$ denotes the charge density of the isolated magnetic tip, and $\rho^{\mathrm{sample}}_\sigma$ denotes the charge density of the adatom adsorbed on the Mn/W(110) surface for spin channel $\sigma$, respectively. 
Here, $\sigma$ is $\uparrow$ for the majority spin channel and $\downarrow$ for
the minority spin channel. Note, that the structural relaxations of the tip
and sample due to their interaction must be taken into account for the
calculations of the isolated systems.
In this way we can reveal the exchange mechanisms 
based on the spin-resolved and spatial change of electron density due to the interaction, i.e., spin-dependent charge accumulation and depletion. 

Figure~\ref{fig:charge_density_Co} shows such spin-resolved CDD plots 
along the [1$\overline{1}$0] direction for interaction of the Fe-terminated tip with the Co adatom adsorbed on Mn/W(110) at a tip adatom distance of $d=0.36$~nm for the $p$- and $ap$-configuration. At this intermediate separation, the electronic interaction between the tip and sample is dominated by the apex atom in accordance with our interpretation of the exchange energies.

If we take a look at the $ap$-conﬁguration (Fig.~\ref{fig:charge_density_Co}(c,d)), we observe that the charge redistribution of
the majority and minority channels are similar to each other
except for an inversion of charge accumulation and depletion
at the tip Fe-apex atom and the interacting Co adatom for both channels. 
This spin-resolved charge redistribution is characteristic for a short-range direct antiferromagnetic exchange mechanism between the $3d$ orbitals due to the formation of spin-dependent covalent bonds~\cite{Williams1981,Tao2009}. 

In the parallel magnetic conﬁguration (Fig.~\ref{fig:charge_density_Co}(a,b)), 
the spin-resolved charge redistributions are quite different. In the majority channel we observe a charge depletion at the Fe apex atom and at the Co adatom and an accumulation 
in the vacuum gap between them. This charge redistribution is the characteristic feature of the indirect Zener-type double exchange mechanism~\cite{Zener1,Zener2,Zener3} which leads to a ferromagnetic exchange coupling between tip and sample~\cite{Tao2009}. The origin of this mechanism is related to the delocalized $sp$-conduction electrons which couple with the $d$ states between the apex and adsorbed surface atoms only for a parallel alignment of the magnetic spin moments due to localized $d$ electrons. In contrast, in the minority channel, the charge density is depleted at the Fe apex atom and accumulated at the probed Co adatom which is the characteristic feature for a direct exchange mechanism between the $d$ orbitals due to the formation of spin-dependent covalent bonds. 

\begin{figure}[htbp!]
	\centering
	\includegraphics[width=0.95\linewidth]{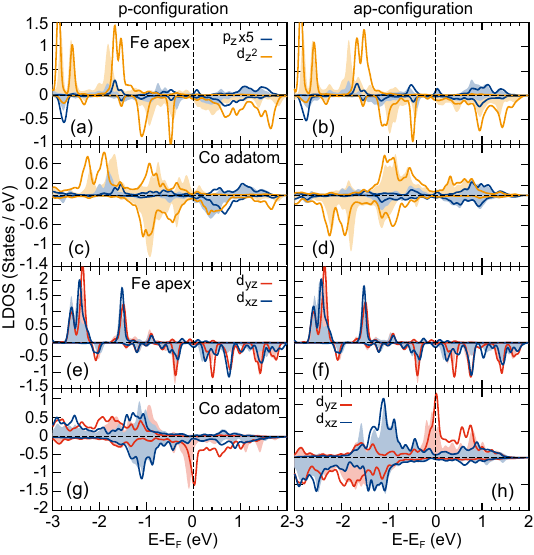}
	\caption{Spin- and orbital-decomposed local density of states (LDOS) for Fe-terminated tips and an adsorbed Co adatom on Mn/W(110) for the $p$- and 
    $ap$-configuration for two different separations $d=0.28$~nm (shaded areas) and $d=0.36$~nm (solid lines). The left and right column shows the $p$- and $ap$-configurations, respectively. The LDOS for spin-up and -down states are
    given on the positive and negative $y$-axis, respectively.
    Panels (a,b) and (c,d) 
    show the spin-resolved 
    LDOS for $p_{\rm z}$ and $d_{{\rm z}^2}$ states
    for the Fe apex and Co adatom, respectively. Note, that the LDOS of the $p_{\rm z}$
    states has been multiplied by a factor of 5. Panels (e,f) and (g,h) 
    show the LDOS for
    $d_{\rm yz}$ and $d_{\rm xz}$ states for the Fe apex and Co adatom, respectively.}
	\label{fig:dos_comparison_Co}
\end{figure}

The analysis of spin- and orbital-decomposed 
local density of states
(LDOS) for the parallel configuration [Fig.~\ref{fig:dos_comparison_Co}(a,c)] indicates a direct hybridization of Fe $p_z$ states with the Co $d_{z^2}$ states. This results in a strong effect in the minority spin channel below the Fermi energy with a larger shift of states to lower energies. However, this is absent in the antiparallel configuration. Instead, one can notice a reduction
of the LDOS in the energy range just below the Fermi energy [Fig.~\ref{fig:dos_comparison_Co}(b,d)] with respect to the parallel configuration [Fig.~\ref{fig:dos_comparison_Co}(a,c)]. 
This effect is clearer at the shorter distance of d=0.28~nm. It is a result of the 
spin-dependent covalent bond formation and favors the antiferromagnetic configuration.
Note, that the changes in the LDOS for the $d_{\rm xz}$ and $d_{\rm yz}$ states is 
smaller for both configurations [Fig.~\ref{fig:dos_comparison_Co}(e-h)].

For the Fe-terminated tip, the Zener-type double exchange mechanism is weaker than the direct $d-d$ antiferromagnetic exchange due to a lower $s-d$ coupling~\cite{Zener1,Zener2,Zener3}, hence we observe an overall antiferromagnetic exchange [cf. Fig.~\ref{fig:SPEX_Compare_on_Co}(a)]. For the tip with a Mn apex atom, we observe antiferromagnetic exchange at small distances for the same reason as well. However, the
indirect ferromagnetic exchange mechanism is much more
long range as it is mediated by the delocalized electrons. 
The larger $s-d$ coupling for the Mn-apex atom results in a stronger double exchange mechanism and hence the total exchange mechanism is ferromagnetic at a larger tip-sample distance [cf.~Fig.~\ref{fig:SPEX_Compare_on_Co}(b)].

\subsection{Mn \& Ir adatom on Mn/W(110)}
\begin{figure}[htbp!]
	\centering
 	\includegraphics[width=0.9\linewidth]{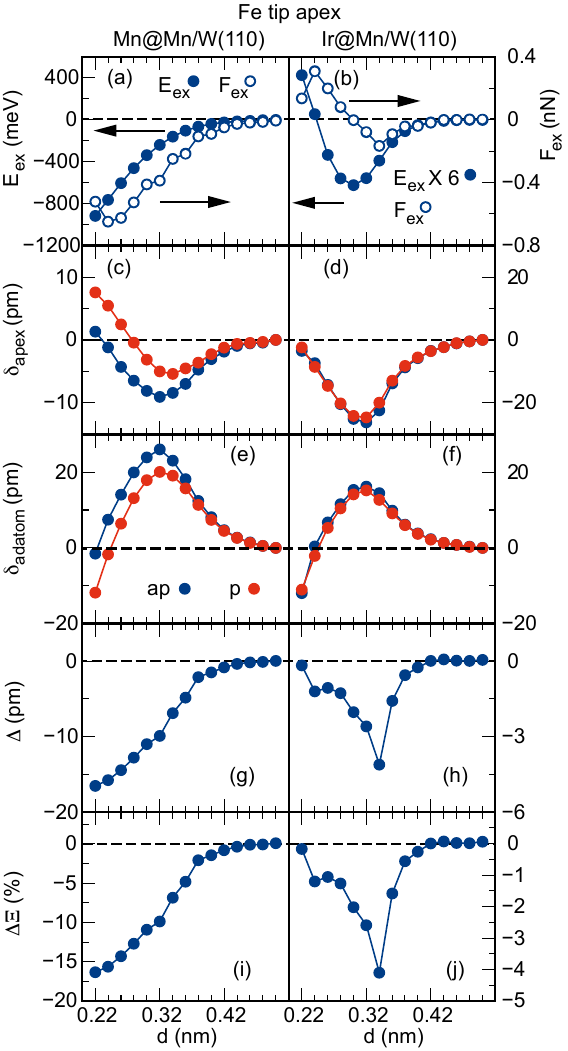}
	\caption{(a,b) Exchange energy (solid dots) and exchange force (hollow circles) as a function of tip-adatom distance $d$ for an Fe tip that approaches a Mn or an Ir adatom on the Mn/W(110) surface, respectively. Structural relaxations of (c,d) the tip apex atom, $\mathrm{\delta_{apex}}$, and (e,f) the adatom, $\mathrm{\delta_{adatom}}$, for 
    $p$ (red dots) and $ap$ (blue dots) alignment of tip and adatom magnetic moments. (g,h) Relaxation difference $\Delta (d)$ between $p$ and $ap$-configuration. (i,j) Calculated TMR difference $\Delta \Xi(d)=\Xi(d)-\Xi_0$. Note the different values
    on the left and right $y$-axis for all panels.}
	\label{fig:SPEX_Compare_IrMn}
\end{figure} 
Now we discuss the magnetic exchange interactions between the magnetic Fe tip and a Mn or
an Ir adatom adsorbed on Mn/W(110). Here we only consider Fe tips with an Fe apex atom. The Mn adatom exhibits a large magnetic
moment of about 3.8~$\mu_{\rm B}$ and ferromagnetic coupling to the nearest-neighbor
Mn surface atoms \cite{Gutzeit2020}. The Ir adatom possesses only a small induced magnetic moment
of about 0.2~$\mu_{\rm B}$ which is aligned parallel to the magnetic moments of the nearest-neighbor Mn
surface atoms \cite{Caffrey2014a}. 

The calculated magnetic exchange energies and forces for the Fe tip probing a
Mn adatom and an Ir adatom are shown in Figs.~\ref{fig:SPEX_Compare_IrMn}(a,b), 
respectively. From the exchange energies we only find a regime of antiferromagnetic exchange coupling (i.e., $E_{\mathrm{ex}} < 0$) for both Mn adatom and Ir adatom. As the tip approaches the adatoms, this coupling increases. The absolute magnitude of the exchange energy is about $\sim$ 4 to 6 times larger for the Mn-adatom 
as compared to the Co or the Ir adatom [cf. Figs.~\ref{fig:SPEX_Compare_on_Co}(a-b) and~\ref{fig:SPEX_Compare_IrMn}(a-b)]. Surprisingly, the exchange energy and 
force for the interaction of the Fe tip with the Ir adatom is of similar magnitude as 
that with the Co adatom.

\begin{figure}[htbp!]
	\centering
	\includegraphics[width=0.8\linewidth]{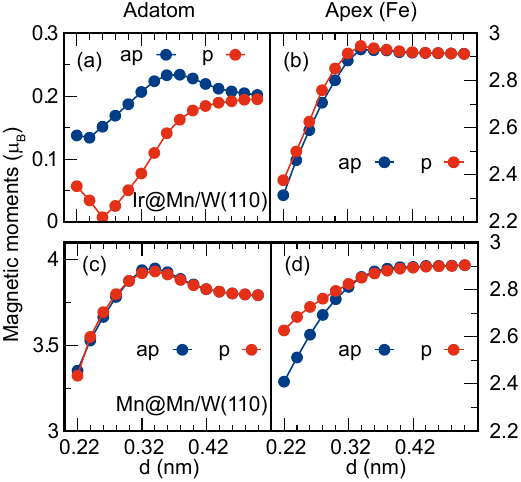}
	\caption{Distance dependence of the magnetic moments for (a,b) the Fe-apex atom and the Ir adatom on Mn/W(110) and (c,d) the Fe-apex atom and the Mn adatom on Mn/W(110). Antiparallel (ap) and parallel (p) configurations are shown in blue and red, respectively. Absolute values of the magnetic moments have been plotted to facilitate comparison between both magnetic alignments.}
	\label{fig:magnetic_moment_IrMn}
\end{figure}

Down to the smallest considered tip-adatom separation there is no local minimum of the exchange energy for the Mn adatom. For the Ir adatom, the exchange energy displays a minimum at a distance of about $d=0.3$~nm. The exchange forces, on
the other hand, show a minimum for both Mn and Ir adatoms. For the Ir adatom
the distance is $\approx 0.34$~nm which could be reached in a scanning
probe experiment. However, for the Mn
adatom the separation is only 0.24~nm which is probably too small in order
to be measured in MExFM before a snap-to-contact occurs between tip and sample. 

The structural relaxations of the Fe apex atom of the tip and of the Mn and
Ir adatom on Mn/W(110) (Fig.~\ref{fig:SPEX_Compare_IrMn}(c-f)) show a similar
qualitative behavior as found for the Co adatom (cf.~Fig.~\ref{fig:SPEX_Compare_on_Co}(c,e)).
However, the Mn adatom displays a much more pronounced effect than the Co adatom with
a maximum displacement of above 20~pm and a large difference between $p$- and 
$ap$-spin alignment (Fig.~\ref{fig:SPEX_Compare_IrMn}(e)). For the Ir adatom the 
displacement reaches a value of almost 20~pm but only a small dependence on the 
magnetic configuration (Fig.~\ref{fig:SPEX_Compare_IrMn}(f)). 
Interestingly, the Fe tip apex atom is much more attracted towards the Ir adatom 
(Fig.~\ref{fig:SPEX_Compare_IrMn}(d))
than to either the Co (cf.~Fig.~\ref{fig:SPEX_Compare_on_Co}(c))
or the Mn adatom (Fig.~\ref{fig:SPEX_Compare_IrMn}(c)). We attribute this
to the larger extent of the $5d$ orbitals of the Ir adatom.

For the interaction of the Fe tip with the Mn adatom on Mn/W(110) we obtain the largest
exchange induced relaxation difference $\Delta(d)$ of all considered systems. The
difference rises continuously starting at about 0.42~nm and reaches values well above 
10~pm (Fig.~\ref{fig:SPEX_Compare_IrMn}(g)). Therefore, we predict a large effect 
for 
the change of the TMR of up to 15\% at the closest distances (Fig.~\ref{fig:SPEX_Compare_IrMn}(i)) that can be measured simultaneously with
the force using the SPEX technique \cite{Hauptmann2017,Hauptmann2019}. In line with
the small dependence of the Fe apex atom and Ir adatom relaxations on the spin configuration,
$\Delta(d)$ is only on the order of a few pm for the case of Ir (Fig.~\ref{fig:SPEX_Compare_IrMn}(h)). Therefore, the TMR changes only by up to 4\%,
however, the drop is quite steep (Fig.~\ref{fig:SPEX_Compare_IrMn}(j)) which might
be visible in an experiment.

\begin{figure}[htbp!]
	\centering
	\includegraphics[width=1.0\linewidth]{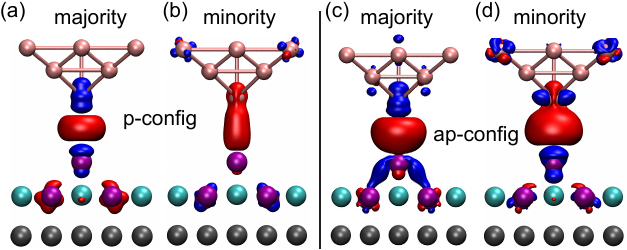}
	\caption{Spin-resolved charge density difference plots along the [1$\overline{1}$0] direction for the interaction of an Fe-tip with a Fe apex atom
    and the Mn adatom on Mn/W(110) at a tip sample distance of $d=0.36$~nm.
    (a,b) Majority and minority spin channel, respectively, for the $p$-alignment
    of the magnetic moments of Fe tip and Mn adatom. (c,d) as (a,b) for the
    $ap$-alignment. Red and blue isosurfaces denotes charge accumulation and depletion, respectively.
   }
	\label{fig:charge_density_Mn}
\end{figure}

In order to elucidate the origin of the exchange energies and forces on the Fe tip
approaching the Ir 
or Mn adatom, we first analyze the magnetic moments and their dependence on distance and
spin alignment between tip and adatom (Fig.~\ref{fig:magnetic_moment_IrMn}). The magnetic
moment of the Fe apex atom remains nearly constant down to relatively small separations of
about 0.32~nm in both cases. At closer distances, we find a significant decrease from its
initial value of about 2.9~$\mu_{\rm B}$ at most 
down to about 2.4~$\mu_{\rm B}$. A small increase
of the Fe moment before the drop is noticeable for the interaction with the Ir adatom 
(Fig.~\ref{fig:magnetic_moment_IrMn}(b)). It occurs for both spin alignments with the
Ir atom and can be explained by the relatively large displacement of the Fe apex atom
which reduces its hybridization with the other tip atoms (Fig.~\ref{fig:SPEX_Compare_IrMn}(d)).
A similar even more pronounced effect is found for the Mn adatom 
(Fig.~\ref{fig:magnetic_moment_IrMn}(c)) and can also be related to the structural change
upon the interaction with the tip (Fig.~\ref{fig:SPEX_Compare_IrMn}(e)). 

For the Ir adatom the curves of magnetic moments look distinctively different (Fig.~\ref{fig:magnetic_moment_IrMn}(a)). The Ir magnetic moment
is very small and only induced by the Mn surface atoms. Upon approaching the Ir adatom with
the Fe tip the magnetic moment increases in the $ap$-alignment, while it decreases for
the $p$-alignment. This different behavior can be attributed to the spin-dependent 
hybridization between Fe apex and Ir atom and can explain the observed exchange
energies and forces (Fig.~\ref{fig:SPEX_Compare_IrMn}(b)). 

\begin{figure}[htbp!]
	\centering
	\includegraphics[width=0.95\linewidth]{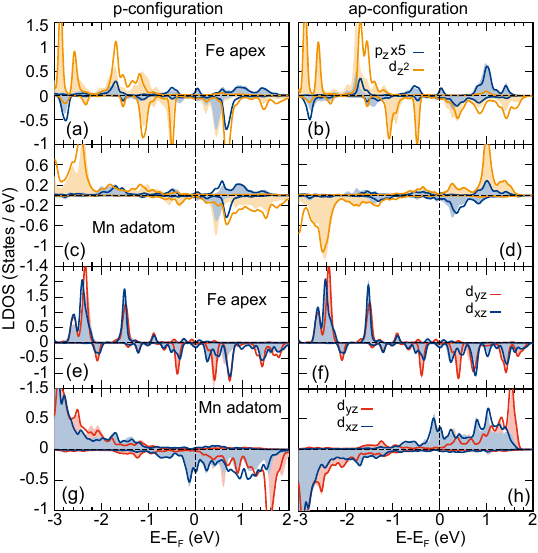}
	\caption{Spin- and orbital-decomposed local density of states (LDOS) for Fe-terminated tips and an adsorbed Mn adatom on Mn/W(110) for the $p$- and 
    $ap$-configuration for two different separations $d=0.30$~nm (shaded areas) and $d=0.36$~nm (solid lines). The left and right column shows the $p$- and $ap$-configurations, respectively. The LDOS for spin-up and -down states are
    given on the positive and negative $y$-axis, respectively.
    Panels (a,b) and (c,d) show the spin-resolved 
    LDOS for $p_{\rm z}$ and $d_{{\rm z}^2}$ states
    for the Fe apex and Mn adatom, respectively. Note, that the LDOS of the $p_{\rm z}$
    states has been multiplied by a factor of 5.
    Panels (e,f) and (g,h) show the LDOS for
    $d_{\rm yz}$ and $d_{\rm xz}$ states for the Fe apex and Mn adatom, respectively. 
     }
	\label{fig:dos_comparison_Mn}
\end{figure}

To gain further insight into the exchange mechanisms for the Mn and Ir adatom we study
spin-resolved charge density difference plots and the local density of states of the
interacting tip-adatom system. For the Mn adatom that is probed by the Fe tip
the CDD plots (Fig.~\ref{fig:charge_density_Mn}) are qualitatively similar to 
that of the Co adatom probed by the same type of tip (Fig.~\ref{fig:charge_density_Co}). 
In particular, we find the Zener-type indirect double exchange mechanism in the majority
channel of the $p$-alignment (Fig.~\ref{fig:charge_density_Mn}(a)) and the 
direct $d-d$ type antiferromagnetic exchange due to covalent bonds
in both spin channels for the $ap$-alignment (Fig.~\ref{fig:charge_density_Mn}(c,d)).

In the spin- and orbital-resolved LDOS for the probed Mn adatom on Mn/W(110) (Fig.~\ref{fig:dos_comparison_Mn}) we observe that there is only a small change
for the $d_{\rm xz}$ and $d_{\rm yz}$ states upon changing the spin-alignment with
the Fe tip from $p$ to $ap$ (Fig.~\ref{fig:dos_comparison_Mn}(e-h)). A 
clearer effect is seen for the orbital types extending
from tip apex atom to the adatom and vice versa, i.e.~the $p_{\rm z}$ and 
$d_{{\rm z}^2}$ states (Fig.~\ref{fig:dos_comparison_Mn}(a-d)).
Upon changing the spin alignment from $p$ to $ap$,
which leads to a switching of spin-up and spin-down LDOS for the Mn adatom
in Fig.~\ref{fig:dos_comparison_Mn}(c) vs.~Fig.~\ref{fig:dos_comparison_Mn}(d), we can see that peaks shift away from
the Fermi energy in the Fe and Mn LDOS due to the direct $d-d$ hybridization. 
This effect is more pronounced at the smaller distance of 0.3~nm 
(shaded areas in the LDOS plots)
and leads to a reduced energy of the $ap$-configuration with respect to
the $p$-configuration that is reflected in the negative exchange energy.

\begin{figure}[htbp!]
	\centering
	\includegraphics[width=1.0\linewidth]{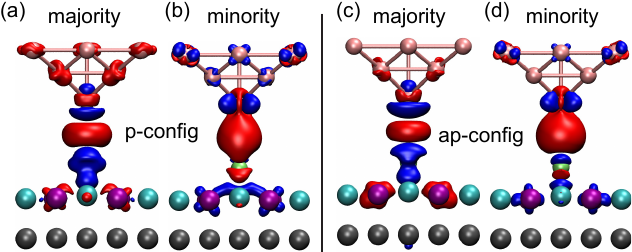}
	\caption{Spin-resolved charge density difference plots along the [1$\overline{1}$0] direction for the interaction of an Fe-tip with a Fe apex atom
    and the Ir adatom on Mn/W(110) at a tip sample distance of $d=0.36$~nm.
    (a,b) Majority and minority spin channel, respectively, for the $p$-alignment
    of the magnetic moments of Fe tip and Ir adatom. (c,d) as (a,b) for the
    $ap$-alignment. Red and blue isosurfaces denotes charge accumulation and depletion, respectively.
    }
	\label{fig:charge_density_Ir}
\end{figure}

The spin-resolved charge density difference plots for the Ir adatom probed by the 
Fe tip are similar in the $p$-configuration (Fig.~\ref{fig:charge_density_Ir}(a,b))
to those
for the Co and Mn adatom. However, in the $ap$-configuration (Fig.~\ref{fig:charge_density_Ir}(c,d)) they do not display the characteristics of
a direct $d-d$ antiferromagnetic exchange mechanism that was visible for the 
$3d$ transition-metal adatoms 
(Fig.~\ref{fig:charge_density_Co} and Fig.~\ref{fig:charge_density_Mn}).  
In particular, in the majority spin channel the charge depletion at the Fe and
Ir atom and the charge accumulation between them is that expected for the
indirect Zener-type ferromagnetic exchange mechanism (Fig.~\ref{fig:charge_density_Ir}(c)). 

\begin{figure}[htbp!]
	\centering
	\includegraphics[width=0.95\linewidth]{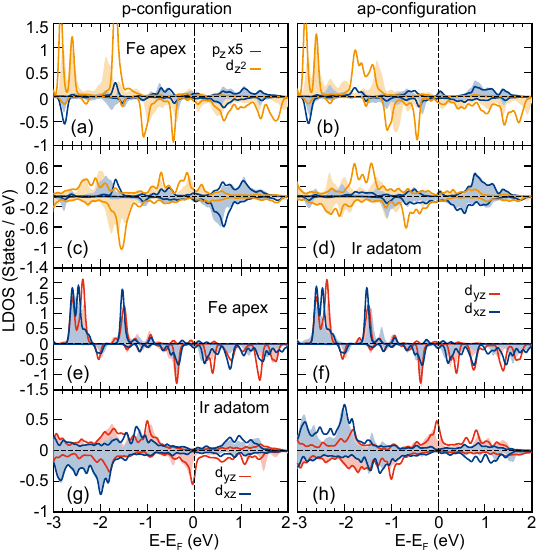}
	\caption{Spin- and orbital-decomposed local density of states (LDOS) for Fe-terminated tips and an adsorbed Ir adatom on Mn/W(110) for the $p$- and 
    $ap$-configuration for two different separations $d=0.30$~nm (shaded areas) and $d=0.36$~nm (solid lines). The left and right column shows the $p$- and $ap$-configurations, respectively. The LDOS for spin-up and -down states are
    given on the positive and negative $y$-axis, respectively.
    Panels (a,b) and (c,d) show the spin-resolved 
    LDOS for $p_{\rm z}$ and $d_{{\rm z}^2}$ states
    for the Fe apex and Ir adatom. Note, that the LDOS of the $p_{\rm z}$
    states has been multiplied by a factor of 5.
    Panels (e,f) and (g,h) show the LDOS for
    $d_{\rm yz}$ and $d_{\rm xz}$ states.
     }
	\label{fig:dos_comparison_Ir}
\end{figure}

Due to the hybridization with the two nearest-neighbor Mn surface atoms
the Ir adatom acquires an induced spin-polarization that is visible in its
LDOS (Fig.~\ref{fig:dos_comparison_Ir}).
In the spin- and orbital-resolved LDOS of the Ir adatom approached by the Fe tip
(Fig.~\ref{fig:dos_comparison_Ir}) we observe that there is little hybridization
between the $d_{\rm xz}$ and $d_{\rm yz}$ orbitals of apex atom and adatom. On
the other hand, a hybridization is visible between $p_{\rm z}$ and
$d_{{\rm z}^2}$ states which depends on the spin alignment between Fe apex atom
and Ir adatom (upper panels of Fig.~\ref{fig:dos_comparison_Ir}). For the $ap$-configuration, the Ir LDOS of $d_{{\rm z}^2}$ states is shifted to lower energies below 
the Fermi energy. As expected this effect is clearer for the smaller tip-adatom 
separation of 0.3~nm. The spin-dependent hybridization between the Fe and Ir atom
which differs for $p$- and $ap$-alignment can explain the observation
of a exchange energy and force with the Fe tip.

\section{Summary and Conclusion}
\label{sec:conclusion}

We have performed a DFT study of the exchange interaction between Fe tips terminated
with an Fe or a Mn apex atom that are approached towards a Co, Mn, or Ir adatom on 
the Mn/W(110) surface.  We have found significant exchange energies on the order 
of up to a few 100~meV and exchange forces of up to few tenth nN
for all three types of adatoms favoring an antiparallel alignment.
The largest values are obtained for the Mn adatom.
Both exchange energy and force curves display a 
characteristic shape with a maximum absolute value at short distances. Since
adatoms can be approached more closely by a tip than a clean surface in atomic 
force microscopy experiments before a snap-to-contact occurs
we anticipate that this local maximum can be revealed 
in future MExFM experiments. 

The exchange mechanisms are analyzed based on spin-resolved charge density difference plots and the local density of states of the interacting tip-adatom system. The origin of the observed antiferromagnetic coupling between tip and adatom is due to a 
short-ranged direct $d-d$ interaction. At larger tip-adatom
separation an indirect long-range Zener-type double exchange mechanism contributes 
which favors ferromagnetic coupling. For the Co adatom approached by a Mn terminated tip
we find a small regime at large separations in which the ferromagnetic coupling
dominates.

Structural relaxations of the tip-adatom junction play a considerable role.
Since the tip-adatom interaction depends on their spin alignment, the actual 
tip-adatom distance differs for the parallel vs.~the antiparallel configuration. 
Therefore, the TMR, which can
be obtained by measuring the tunneling current in SP-STM or SPEX, increases in
absolute value as the tip approaches the adatom. This TMR change can be from a few 
\% for a Co adatom up to 15\% for a Mn adatom and depends also on the tip termination.

The results obtained in our work provide insight into the
exchange interactions at the single-atom level and can be valuable to trigger and
guide future experiments using MExFM or SPEX imaging on magnetic adatoms.

\section*{Acknowledgments}
It is our pleasure to thank Alexander Khajetoorians and Nadine Hauptmann for valuable discussions. 
We gratefully acknowledge financial support from the Deutsche Forschungsgemeinschaft (DFG, German Research Foundation) via Project 
No.~445697818 and computing time provided by the North-German Supercomputing Alliance (HLRN).

\end{document}